\documentclass[12pt]{article}

\usepackage{graphicx}

\usepackage{scicite}

\usepackage{times}

\newenvironment{sciabstract}{%
\begin{quote} \bf}
{\end{quote}}

\newcounter{lastnote}
\newenvironment{scilastnote}{%
\setcounter{lastnote}{\value{enumiv}}%
\addtocounter{lastnote}{+1}%
\begin{list}%
{\arabic{lastnote}.}
{\setlength{\leftmargin}{.22in}}
{\setlength{\labelsep}{.5em}}}
{\end{list}}

\title{Fast optical variability of Naked-Eye Burst -- manifestation of periodic
  activity of internal engine.}

\author
{G.~Beskin$^{1\ast}$, S.~Karpov$^1$, S.~Bondar$^2$, \\
  A.~Guarnieri$^3$, C.~Bartolini$^3$, G.~Greco$^3$, A.~Piccioni$^3$ \\
  \\
  \normalsize{$^1$Special Astrophysical Observatory, Nizhniy Arkhyz,
    Karachai-Cirkassia, Russia,} \\
  \normalsize{$^2$Institute for Precise Instrumentation, Nizhniy Arkhyz,
    Karachai-Cirkassia, Russia,} \\
  \normalsize{$^3$Astronomical Department of Bologna University, Bologna, Italy,} \\
\\
\normalsize{$^\ast$To whom correspondence should be addressed; E-mail:  beskin@sao.ru.}
}

\date{}

\begin{document} 


\baselineskip24pt


\maketitle


\begin{sciabstract}
  We imaged the position of the Naked-Eye Burst, GRB080319B, before, during and
  after its gamma-ray activity with sub-second temporal resolution and
  discovered the fast variability of its prompt optical emission. Its
  characteristics and similarity with properties of gamma emission temporal
  structure suggest that it reflects the behaviour of internal engine --
  supposedly, a hyperaccreting solar-mass black hole formed in the collapse of
  a massive stellar core.
\end{sciabstract}

% In setting up this template for *Science* papers, we've used both
% the \section* command and the \paragraph* command for topical
% divisions.  Which you use will of course depend on the type of paper
% you're writing.  Review Articles tend to have displayed headings, for
% which \section* is more appropriate; Research Articles, when they have
% formal topical divisions at all, tend to signal them with bold text
% that runs into the paragraph, for which \paragraph* is the right
% choice.  Either way, use the asterisk (*) modifier, as shown, to

%% \section*{Introduction}

In the last several years a general picture that explains the basic observed
properties of gamma-ray bursts has been developed \cite{piran, meszaros}. Its
main part is the formation of a compact relativistic object, a black hole or a
highly-magnetized neutron star, as the result of either the collapse of the
core of a massive star\cite{woosley, paczynski}, or the merging of two neutron
stars\cite{eichler_merging}. Such a black hole is surrounded by a massive disk
of residual matter that avoided the collapse; and its accretion and interaction
with the envelope leads to the launch of relativistic ejecta; its kinetic
energy eventually transforms to the emission of various wavelengths. Internal
instabilities or shocks produces the gamma-ray burst itself, while the
subsequent interaction of the ejecta with interstellar medium manifests as a
lower energy afterglow.  Temporal structure of such flares
%, their light curves,
reflects both the behaviour of the ``internal engine'' -- accreting black hole or
magnetar -- and the structure and dynamics of the ejecta. Activity of
``internal engine'' is supposedly periodical in any model, while the behaviour
of the ejecta and their emission is rather stochastic and is determined by the
mechanisms of conversion of kinetic energy to an internal (thermal and
magnetic) one of emitting particles, by the emission mechanism itself and by
the structure of the
ejecta\cite{kobayashi,daigne_moskovitch,nakar_piran,lyutikov}.

Detailed long-term study of gamma-ray burst light curves have not yet succeeded
in discovery of neither periodic variations nor any other direct signatures of
``internal engine'' behaviour.

Detection of optical flashes accompanying the gamma-ray bursts and study of
their light curves may provide significantly new information on their
physics. Indeed, discovery of prompt optical emission in follow-up observations
of ROTSE-I\cite{prompt_akerlof}, RAPTOR\cite{prompt_raptor},
ROTSE-III\cite{prompt_rotse3} and UVOT\cite{prompt_uvot} have lead to
refinements of different GRB model details. However, it has not clarified the
nature of the bursts well enough, mainly due to insufficient temporal
resolution of the observations (typically worse than 10 s), that prevents the
analysis of detailed structure of optical light curves and their relation to
gamma-ray ones.

As a result of generic stategy of the searching for and investigation of fast
optical flashes accompanying gamma-ray bursts\cite{piccioni,beskin_scanning} we created the
FAVOR\cite{beskin_favor2} and TORTORA\cite{molinari_tortora} wide-field optical
cameras with subsecond temporal resolution. The latter has succeeded in
observing
%discovery and detailed study of
the Naked-Eye Burst\cite{naked_karpov,naked_nature,naked_china}.
A detailed presentation and analysis of these observations are the aim of this paper.

%% \section*{Observations}

\begin{figure}[t]
  {\centering \resizebox*{\columnwidth}{!}{\includegraphics[angle=0]{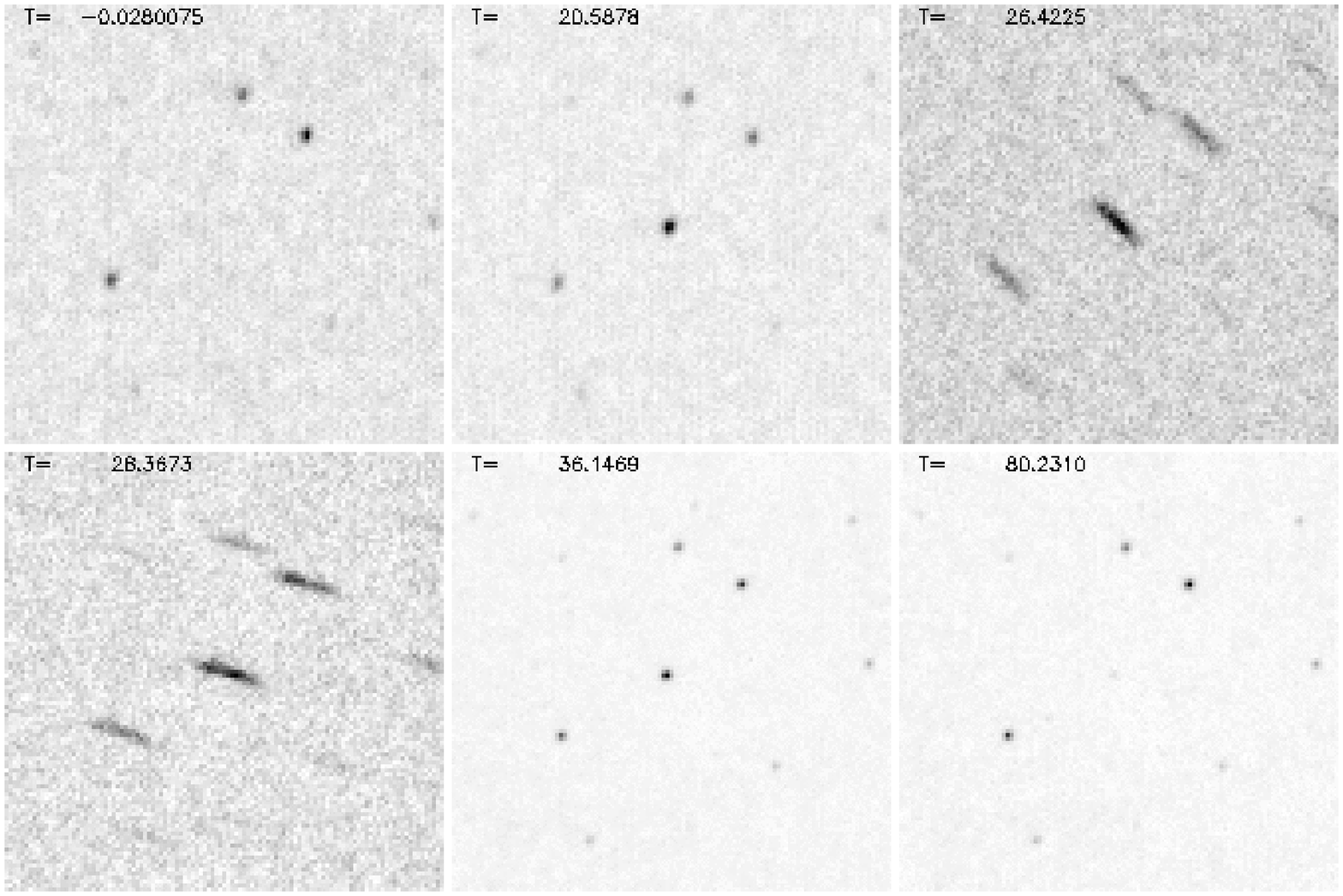}}}
  \caption{The development of prompt optical emission from GRB080319b as seen
    by TORTORA camera. Sums of 10 consecutive frames with 1.3 s effective
    exposures are shown for the gamma-ray trigger time ($T=0$ s), the maximum
    brightness time during the first peak ($T=20.5$ s), two middle-part moments
    ($T=26.4$ s and $T=28.4$ s), at the last peak ($T=36$ s) and during early
    afterglow ($T=80$ s) stages. Image size is 2.5 x 2.5 degrees. The third and
    fourth images display deformed star profiles as during this time (since
    $T+24$ s till $T+31$ s) REM robotic telescope (which has TORTORA camera
    mounted on top) repointed after receiving the burst information from
    Swift. Initially, burst position was on the edge of field of view, as a
    result of repointing it moved to the center of field of view, which
    resulted in better data quality.}
  \label{fig_naked_images}
\end{figure}

% \begin{figure}[t]
%   {\centering \resizebox*{\columnwidth}{!}{\includegraphics[angle=270]{naked_mags.eps}}}
%   \caption{The light curve of GRB080319B acquired by TORTORA wide-field
%     camera (upper curve, left axis) alongside with Swift BAT gamma-ray one
%     (lower curve, right axis). Also, transient brightness measurements by ``Pi
%     of the Sky'' optical camera\cite{naked_nature} are shown.
%     Swift-BAT light curve is a sum of all four energy channels. TORTORA data
%     points show both full-resolution (original data frames) and
%     low-resolution (10 images co-added). Full-resolution data are unavailable
%     for a period of REM telescope repointing due to massive blurring of object
%     image. Only data points with errors less than the value are shown.
%   }
%   \label{fig_naked_mags} 
% \end{figure}

\begin{figure}[t]
  {\centering \resizebox*{\columnwidth}{!}{\includegraphics[angle=270]{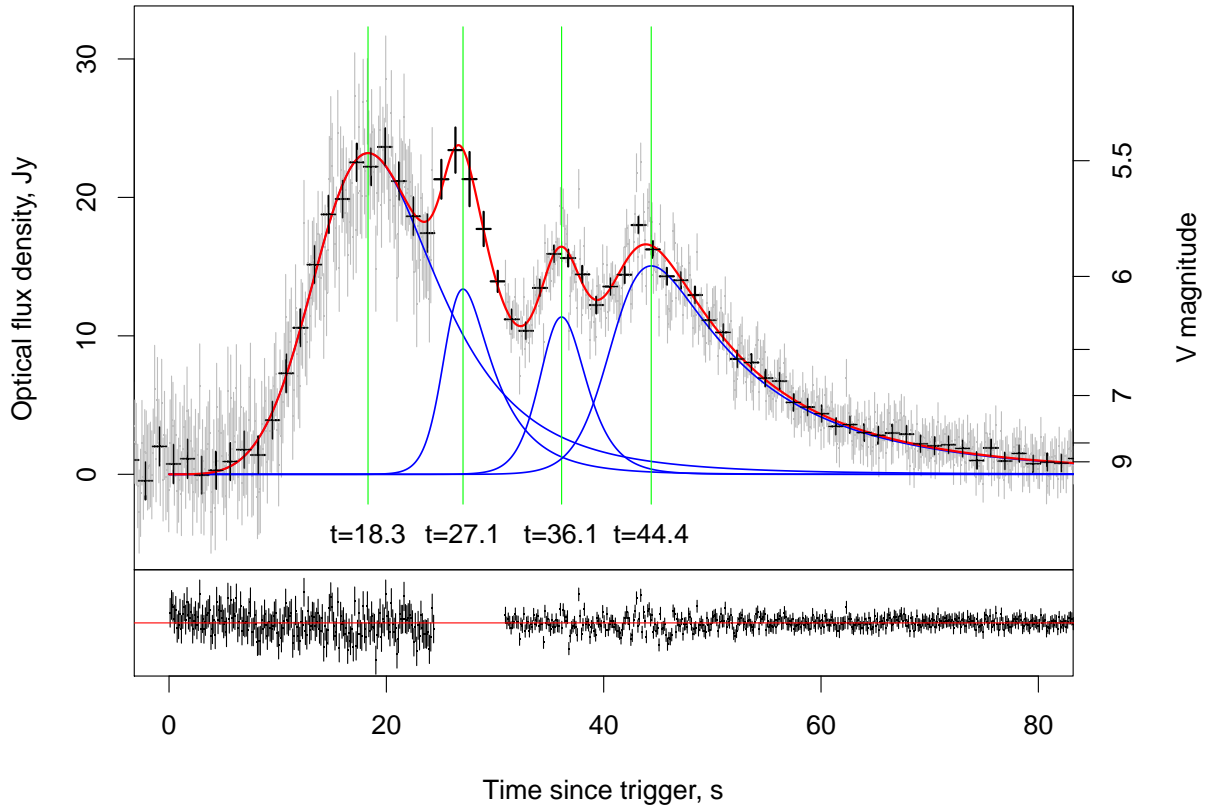}}}
  \caption{The light curve of GRB080319B acquired by TORTORA wide-field
    camera. The gamma-emission started at $T\approx-4$ s and faded at
    $T\approx55$ s. Full resolution (0.13s exposure, gray lines) data are
    available for all duration of gamma-emission except for interval of REM
    telescope repointing ($24.5$ s $ < T < 31$ s), while low-resolution ones
    (summation of 10 consecutive frames, 1.3 s effective exposure) -- for the
    whole time. The light curve is approximated by a four nearly equidistant
    flares; lower panel shows the residuals of such approximation.}
  \label{fig_naked_lc} 
\end{figure}

\begin{table}
  \centering
  \begin{tabular}{c|c|c|c|c}
    \hline
    $T_0, s$ & $F_0, Jy$ & $r$ & $d$ & $\Delta T, s$ \\
    \hline
    18.3 $\pm$ 0.3 & 23.2 $\pm$ 0.6 & 4.0 $\pm$ 0.4 & -5.4 $\pm$ 4.1 & 8.7$\pm$ 0.4\\
    27.0 $\pm$ 0.3 & 13.4 $\pm$ 3.4 & 24.8 $\pm$ 8.3 & -9.7 $\pm$ 4.9 & 9.1$\pm$ 0.4 \\
    36.1 $\pm$ 0.2 & 11.4 $\pm$ 1.7 & 25.9 $\pm$ 7.6 & -22.0 $\pm$ 17 & 8.3$\pm$ 0.5 \\
    44.4 $\pm$ 0.5 & 15.1 $\pm$ 1.8 & 21.9 $\pm$ 3.3 & -5.1 $\pm$ 0.2 &  \\
    \hline
  \end{tabular}
  \caption{Best-fit parameters for the decomposition of the light curve into 4
    peaks with shape described by Kocevski \cite{kocevski_2003} profile, shown
    in Figure~\ref{fig_naked_lc}. Here,
    $T_0$ and $F_0$ are the peak maximum positions relative to trigger time
    and fluxes, while $r$ and $d$
    are the power-law indices of their rising and declining parts. $\Delta T$
    is the distance between the peak and the next one.}
  \label{table_decomposition}
\end{table}

\begin{figure}[t]
  {\centering \resizebox*{\columnwidth}{!}{\includegraphics[angle=270]{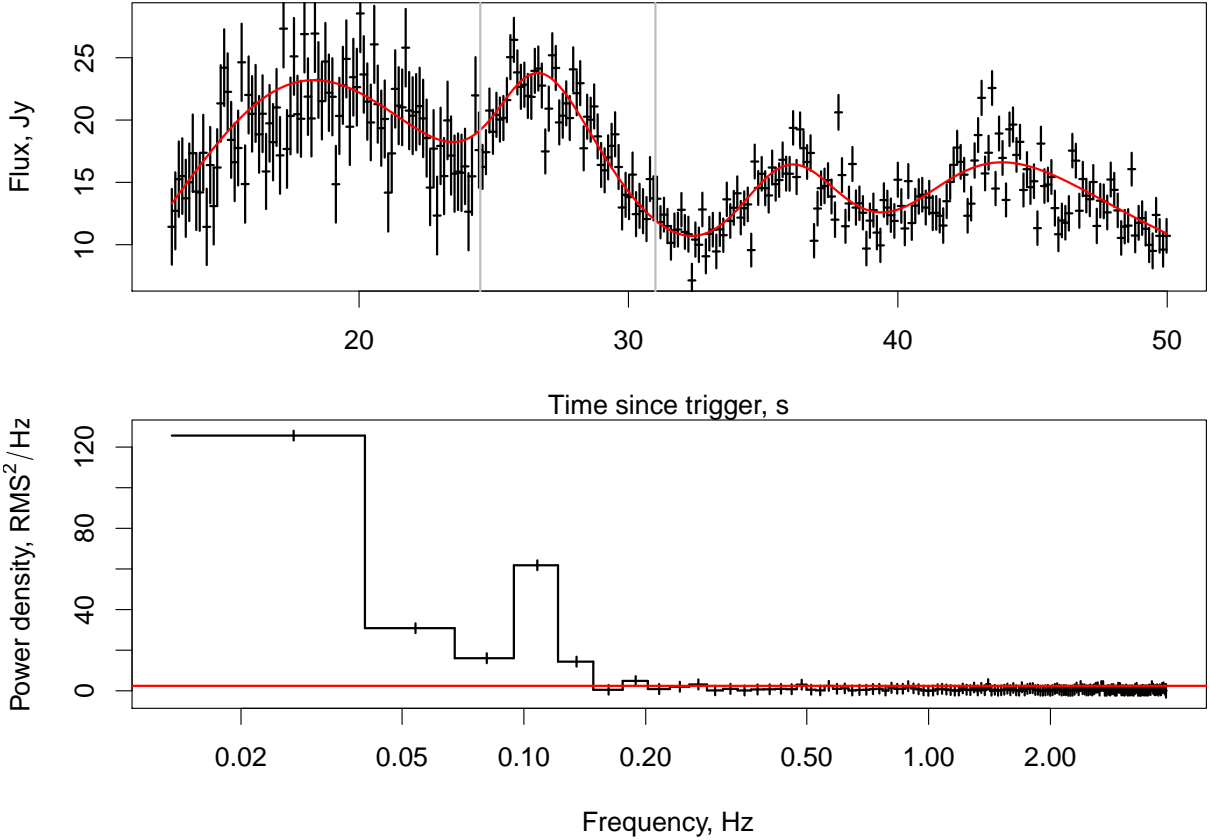}}}
  \caption{Upper panel -- the plateau stage of the burst optical emission from
    $T+13$ s till $T+50$ s. Smooth line shows the superposition of four peaks
    with parameters given it Table~\ref{table_decomposition}.  High-resolution
    data for a missing part (since $T+24.5$ s till $T+31$ s, marked with gray
    vertical lines) has been simulated by a white noise with a mean
    corresponding to the smooth curve and a standard deviation corresponding to
    the one around the gap.  Lower panel -- power density spectrum of the
    plateau stage. Mean noise level (horizontal line) and error bars are
    estimated by bootstrapping method -- by generating a large number of sample
    time series by randomly shuffling the original light curve, what completely
    destroys its time structure while keeping the distibution of its values,
    and by studying the distribution and quantiles of resulting power
    densities. The feature at $\sim$9 s is clearly visible with significance
    level better than at least $10^-7$, and corresponds to four
    nearly-equidistant peaks of the light curve. Low-frequency (less than 0.05 Hz)
    excess reflects the difference of mean intensity levels of first and second
    halves of the light curve.}
  \label{fig_naked_power} 
\end{figure}

\begin{figure}[t]
  {\centering \resizebox*{\columnwidth}{!}{\includegraphics[angle=270]{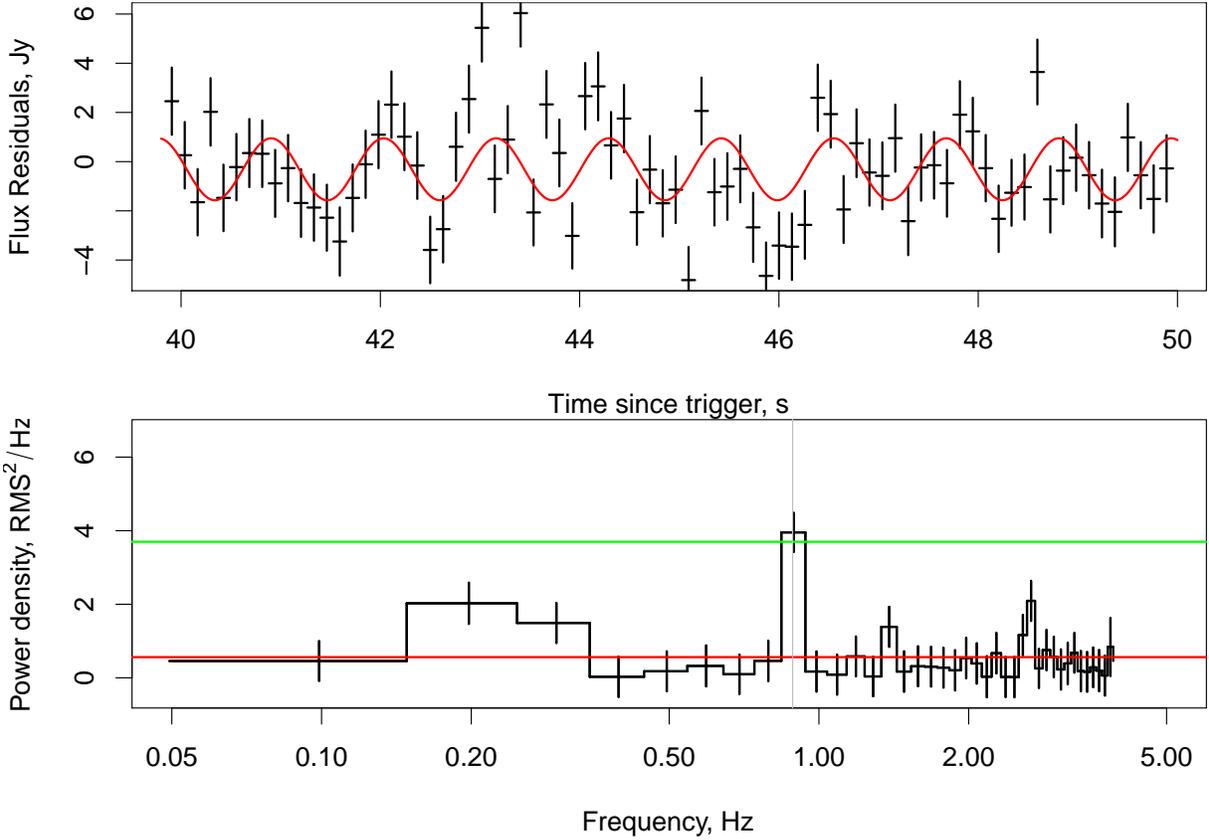}}}
  \caption{Upper panel -- optical flux for a $T+40$ s -- $T+50$ s interval
    (last peak) with the approximation shown in Figure~\ref{fig_naked_lc}
    subtracted.  Smooth line shows the best-fit sinusoidal approximation of the
    data with $P=1.13$ s period.  Lower panel -- power density spectrum of this
    data, estimated the same way as described in Figure~\ref{fig_naked_power}
    caption. Horizontal lines represent mean noise level (lower) and a level of
    noise deviations with $10^{-3}$ significance (upper), estimated by
    bootstrapping number of time series from the original data set. Vertical
    line corresponds to the period of the sinusoidal approximation shown in
    upper panel, clearly coincided with the peak of power spectrum. The
    probability of a random appearance of a feature like the one seen in any of
    39 frequency bin is $\sim0.01$.}
  \label{fig_naked_power_lastpeak}
\end{figure}

\begin{figure}[t]
  {\centering \resizebox*{\columnwidth}{!}{\includegraphics[angle=270]{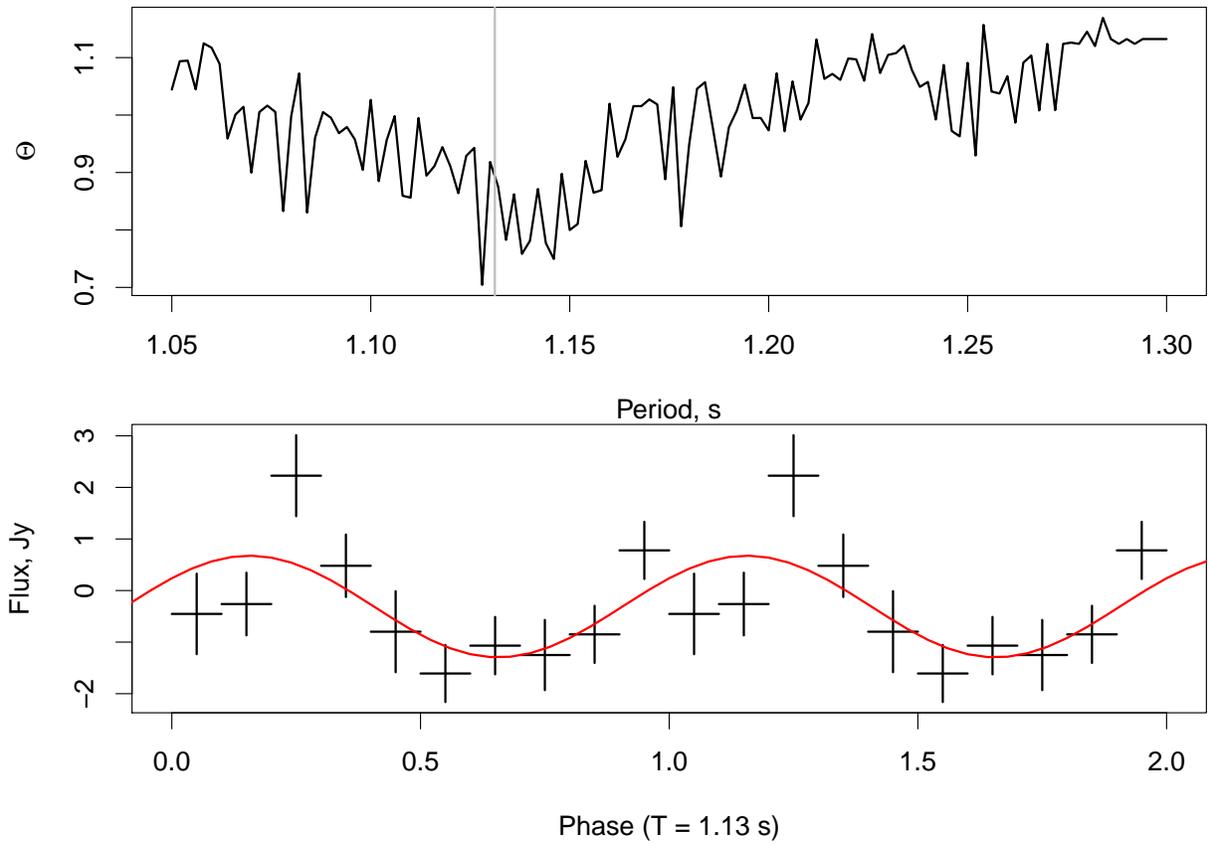}}}
  \caption{Upper panel -- Phase Dispersion Minimization
    \cite{stellingwerf_1978} $\Theta$ statistic for best folding
    period. Gray vertical line represent the $P=1.13$ s period from sinusoidal
    best-fit shown in Figure~\ref{fig_naked_power_lastpeak}. Bottom panel --
    folded mean light curve corresponding to this period, and the sine profile
    corresponding to the mentioned best-fit.}
  \label{fig_naked_fold_lastpeak}
\end{figure}

\begin{figure}[t]
  {\centering \resizebox*{\columnwidth}{!}{\includegraphics[angle=270]{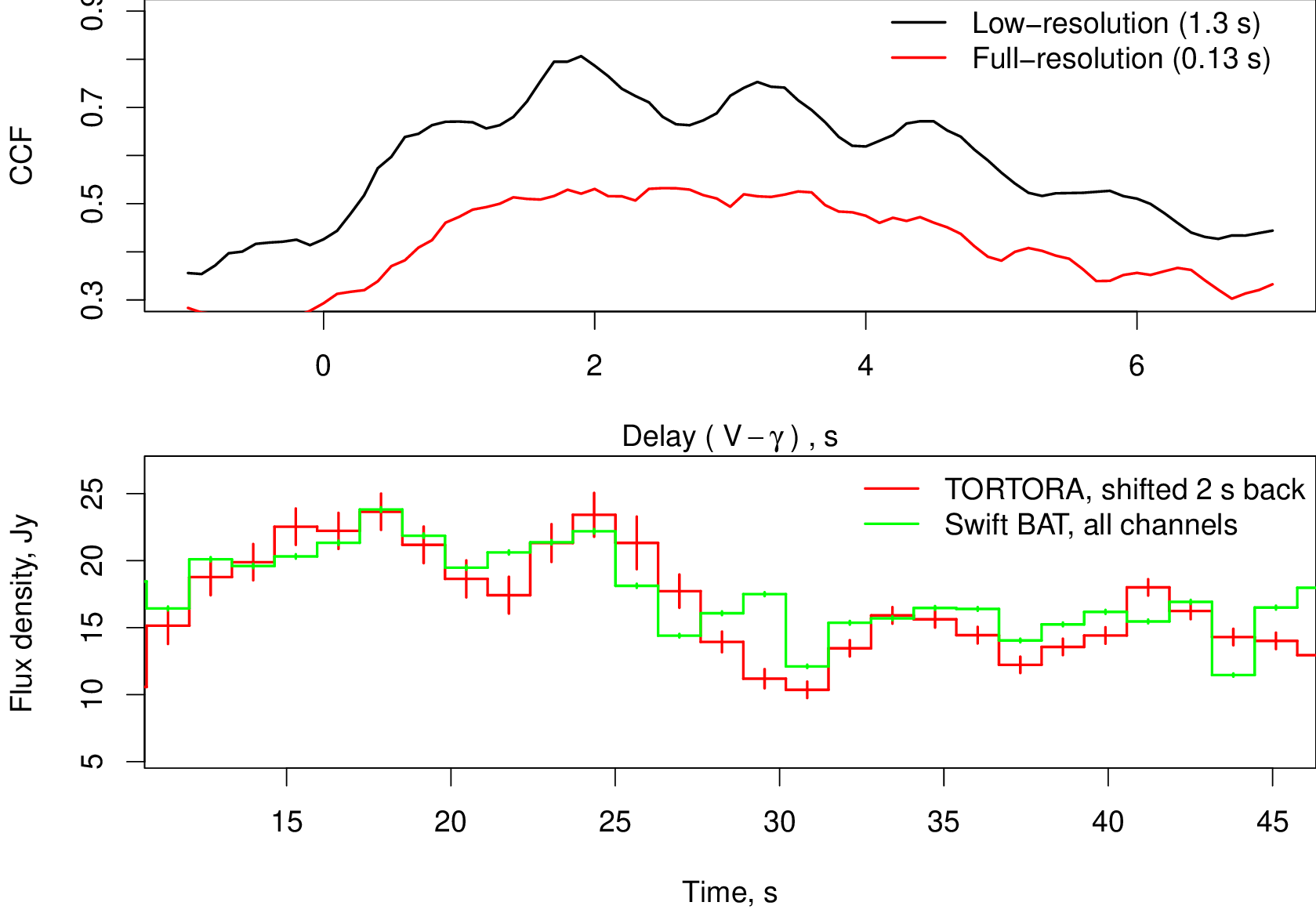}}}
  \caption{Upper panel -- cross-correlation of the Swift-BAT gamma-ray (all
    energy channels) and TORTORA optical fluxes for the main (plateau) phase of
    the burst emission. Lower panel -- TORTORA optical flux shifted back 2
    seconds along with correspondingly rebinned Swift-BAT gamma-ray flux. The
    correlation is $r=0.82$ with significance level of $5\cdot10^{-7}$. Gamma-ray
    curve is arbitrarily scaled and shifted for illustrative purposes.}
  \label{fig_naked_correlation}
\end{figure}

TORTORA is a small (12 cm objective diameter) telescope equipped with image
intensifier and a fast (7.5 s$^{-1}$ frame rate, 0.13 s exposures with no gaps)
TV-CCD camera which routinely monitors 24$^{\circ}$x30$^{\circ}$ area of the
sky, covering part of current Swift gamma-ray telescope field of view to catch the
initial stages of gamma-ray bursts independently from gamma-ray satellite
triggers. It is mounted on top of the Italian REM telescope\cite{molinari_tortora}
at La-Silla observatory (ESO, Chile).

We observed the region of GRB080319B\cite{naked_karpov,naked_nature} since 05:46:22 UT,
nearly half an hour before the burst (burst time is 06:12:49 UT), during the
event and for several tens of minutes after its end. Since 06:13:13 UT till
06:13:20 UT REM telescope performed automatic repointing after receiving the
coordinates distributed by Swift\cite{naked_racusin}, which moved the position
of the burst from the edge of the TORTORA field of view towards its center. Sample
images of the burst region at different stages of the event are presented in
Figure~\ref{fig_naked_images}.

% Figure~\ref{fig_naked} shows the sample 2.5 x 2.5 degrees images centered at
% burst position for its different phases. TORTORA limiting magnitude in that
% conditions had been significantly lower than ones of ``Pi of the Sky'' and
% RAPTOR, but its superior time resolution allowed to trace the burst time
% structure with unprecedented level of details (see Fig.~\ref{fig_naked_lc} for
% summary light curves of TORTORA, ``Pi of the Sky'' and Swift BAT).

The data acquired have been processed by a pipeline including TV-CCD noise
subtraction, flat-fielding to compensate vignetting due to objective design,
and custom aperture photometry code taking into account non-Poissonian and
non-ergodic pixel statistics caused by image intensifier. For the REM
repointing time interval fluxes have been derived using custom elliptic
aperture photometry code after summation of 10 consecutive frames (1.3 s
effective exposure) with compensated motion of the stars, therefore no
full-resolution measurements (0.13 s exposure) is available for this interval.

% Unfortunately, it seems impossible to reliably
% reconstruct the light curve of this interval with any better resolution due to
% massive blurring of star PSF caused by their motion.  For all other intervals,
% photometry has been performed both with 10-frames (1.3 s effective exposure)
% binning, and with original (0.13s) time resolution.

TORTORA acquired the data in white light with sensitivity defined by the S20
photocathode used in the image intensifier\cite{beskin_favor2}. Instrumental
object magnitudes have then been calibrated to Johnson V system using several
nearby Tycho-2\cite{tycho2} stars. A quick-look low resolution light curve
(lacking the data during the REM repointing interval) has been published
\cite{naked_karpov,naked_nature} and has been found to agree with results of
other wide-field monitoring cameras which also observed this burst, such as
``Pi of the sky''\cite{naked_pi} and RAPTOR\cite{naked_raptor}. Our complete
full resolution light curve along with the low resolution one (after the
restoring of the gap) are shown in Figure~\ref{fig_naked_lc}.

We clearly detected the transient optical emission since approximately 10
seconds after the trigger. It then displayed fast $\sim t^{4}$ rise, peaked at
$V\approx5.5^{\rm m}$, demonstrated 1.5-2 times variations on a several seconds
time scale and decayed as $\sim t^{-4.6}$ until went below TORTORA detection
limit at about hundred seconds since trigger. The gamma emission itself ended
at 57th second.

The light curve clearly shows four peaks with similar amplitudes, durations and
shapes. We decomposed it into four components described by a simple Kocevski
\cite{kocevski_2003} profile whose parameters are given in
Table~\ref{table_decomposition}. We stress that distances between peaks are
nearly the same within the errors, and are around 8.5 s in observer frame,
which corresponds to 4.4 s in the rest frame at
z=0.937\cite{naked_nature}. Power spectral analysis of the plateau stage,
excluding the rise and decay parts, also clearly reveals the separate peak at
frequencies around this time scale (see Figure~\ref{fig_naked_power}).

Therefore, for the first time, we have a clear detection of a periodic
variations of prompt optical emission on a few seconds time scale.

We then subtracted the smooth curve, formed by four fitted peaks, from the
original data and studied the residuals shown in the lower panel of
Figure~\ref{fig_naked_lc}. Power spectral analysis of differerent sub-intervals
of the burst revealed the signature of a periodic intensity variations during
the last peak, since $T+40$ s till $T+50$ s, shown in Figure
\ref{fig_naked_power_lastpeak}. No other intervals of the light curve show any
variability in 0.1-3.5 Hz (0.3-10 s) range with power exceeding 15\% before and
10\% after the REM repointing. To exclude artificial nature of these variations
we performed analysis of each comparison star separately in the same way as of
the object.  Neither comparison stars nor background display any similar
periodic feature during either the whole time interval or the last peak.

The significance level of the power density spectrum feature shown in
Figure~\ref{fig_naked_power_lastpeak} is approximately 1\%.  The period and
amplitude of the corresponding sinusoidal component, derived by means of
non-linear least squares fit, are 1.13 s and 9\%, respectively.  The Phase
Dispersion Minimization\cite{stellingwerf_1978} method applied to the data (see
Figure~\ref{fig_naked_fold_lastpeak}) also reveals the signature of a period
near this time; the folded data show quasi-sinusoidal broad profile.

Such one-second periodicity is the fastest optical variability known to date
for any extragalactic object.

To compare the temporal structure of optical and gamma-ray light curves we
performed the cross-correlation analysis, using the plateau phase only,
excluding the first and last 12 seconds of the burst both in optical and in
gamma, which are obviously highly correlated\cite{naked_china} (see
Figure~\ref{fig_naked_correlation}).
The correlation between the full-resolution optical
data and the correspondingly rebinned gamma-ray one is no more
than 0.5, due to high level of stochastic component in 0.1-1 s range in both
optical (measurements noise) and gamma rays (actual high-frequency variability)
\cite{margutti}. For the low-resolution data, with a 1.3 s binning, the
correlation coefficient is, however, as high as 0.82 if the optical light curve
is shifted 2 seconds back with respect to gamma-ray one (see
Figure~\ref{fig_naked_correlation}). Correspondingly rebinned gamma-ray data
demonstrate the same four nearly equidistant peaks as optical ones.

This is the first detection of a close relation between the temporal structures
of the optical and gamma-ray prompt emission. In our the case, the
gamma-ray burst itself precedes the optical flash by two seconds. This result is
for a period of main energy release, while the highly correlated phases of
emission rise and decay have different shapes in different energy ranges.

The $\Delta t \sim 2$ s delay of optical flash relative to gamma-ray one
inevitably suggests that they were generated in different parts of the
ejecta, and optical photons came from the distance
${\Delta R \approx 2c\Gamma^2\Delta t (1+z)^{-1} = 5.4\cdot10^{15} \Gamma_{300}^2}$
farther from the central engine, where $\Gamma_{300}$ is the Lorentz factor in
units of 300 \cite{piran,li_waxman}.

The peculiarities we detected in Naked-Eye Burst clearly contradict models of
the emission generation based on various kinds of interactions between a single
ensemble of electrons and photons they generate (synchrotron or inverse Compton mechanisms)
\cite{naked_nature,kumar_panaitescu,fan_piran}. The same can be said for a
model with two shock waves where the internal forward one produces the optical
emission and the internal reverse one -- gamma photons \cite{yu_wang_dai}, or for
a relativistic turbulence model \cite{narayan_kumar,kumar_narayan}.  On the
other hand, the fast rise and similarity of durations of all four optical
flashes rule out an external shock (both forward and reverse) as a source of
optical emission \cite{zou_piran_sari}.

At the same time, the temporal shift of the light curves, the absence of
optical flux variations on 0.1-1 s in contrast to significant stochastic
variability of the gamma one on the same time scale \cite{margutti}, and the
strong (by thousand times) excess of optical flux spectral density in
comparison to the gamma-ray one \cite{naked_nature} may be explained in the
internal shock with residual collisions model \cite{li_waxman}. In this model
both high energy and optical photons are produced by a synchrotron emission of
electrons on different distances from the central engine -- the closer the
harder the emission. The finer structure of high-energy emission is defined by
the dynamics of collisions of shells with different Lorentz factors, whose
kinetic energy heats the electrons by means of the shock waves. The number of
colliding shells decreases with the distance (as shells merge in a collisions),
which effectively smooths the optical light curve \cite{li_waxman}. Such a
model, obviously, has its own difficulties
\cite{kumar_mcmahon,narayan_kumar,kumar_narayan,zou_piran_sari}, but it
explains the whole set of observational data with a reasonable values of a bulk
Lorentz factor $\Gamma\sim300$, total energy $E\sim10^{54}-10^{55}$ erg,
interstellar medium density $n\sim0.1-1$ cm$^{-3}$ and the emitting electrons
Lorentz factor $\gamma_e\sim100$ \cite{li_waxman,zou_piran_sari}. The gamma-ray
emission is then generated at a distance
${R_{\gamma} < 2c\Gamma^2\tau_{\gamma}(1+z)^{-1} =
  2.7\cdot10^{14}\Gamma_{300}^2}$
cm \cite{piran_nakar}, where the characteristic time scale of gamma-ray
variability is ${\tau_{\gamma} \sim 0.1}$ s \cite{margutti}, while the optical
one -- on a ${R_{opt}\sim10^{16}}$ cm distance. The latter is a consequence of
obvious inequality ${\Delta R < R_{opt} < 2c\Gamma^2\tau_{opt}(1+z)^{-1}}$ or
${5.4\cdot10^{15} < R_{opt} < 1.4\cdot10^{16}}$ cm for a characteristic optical
variability time scale $\tau_{opt}\sim4-5$ s (observed peaks rising time) and
$\Gamma\sim300$.

It is worth noting that the conclusion of generation of optical and gamma
emission of Naked-Eye Burst at different distances from the central engine is a
direct consequence of the detected shift of optical light curve with respect to
gamma-ray one. It does not depend on particular mechanisms of conversion of
mechanical energy to internal one of electrons nor the emission mechanisms of
the latters. Taking into account the high level of similarity of optical and
gamma variability on 8-9 s time scale (see Figure~\ref{fig_naked_correlation})
we inevitably conclude that these variations have the same cause -- namely,
the cyclic variations of internal engine activity (each flash of the light curve
corresponds to one of four its episodes). Obviously, it would be
impossible for the relativistic ejecta itself to display similar
dynamics, both spatial and temporal, in regions separated by $10^{16}$ cm.

Therefore, we can concluse that this is the first time in which the signature
of non-stationary physical processes related to the internal engine of a
gamma-ray burst was discovered in temporal structure of its prompt emission.

These cyclic variations may be the signatures of a black hole formed in a
collapse of a progenitor star, surrounded by a massive hyperaccreting disk and
launching the relativistic ejecta which produce the gamma-ray burst emission
\cite{woosley,macfadyen_woosley,zhang_woosley_heger}. The non-stationarity of
the ejection flow is a result of non-stationary accretion -- the cyclic
increase (by several orders of magnitude) of the accretion rate -- due to
gravitational instability \cite{masada} in the hot inner part of the disk. The
matter accumulates there due to suppression of the magnetorotational
instability, driving the angular momentum transport, due to high neutrino
viscosity in the inner parts of the disk. In contrast, in outer, neutrino
transparent parts of the disk the development of magnetorotational instability
provides ``normal'' accretion rate. When sufficiently high amount of matter
accumulates, the gravitational instability develops in inner parts of the disk,
which significantly increase the local accretion rate. Then, after depletion of
matter, the inner parts of the disk stabilize again. Such a process repeats
until all the matter of the disk fall towards the black hole. As a result, the
ejecta are modulated on a characteristic time scale of the matter accumulation
and instability. On shorter scales, ejecta consist of separate blobs formed
as a result of gravitational instability, whose collisions may result in an
internal shocks.

Without going into a detailed discussion of such a model \cite{masada}, we just
note that nearly all observational properties of Naked-Eye Burst -- duration of
$\sim$50 s, ejecta energy of $10^{54}-10^{55}$ erg, characteristic variability time
scale in the source frame of $\sim$2.5 s -- may be produced by an accretion of a
one solar mass accretion disk with inner zone radius of 300 km (30 gravitational
radii) onto a 3 solar masses black hole.  Moreover, for the same set of
parameters the $\sim$0.5 s variations in source frame, seen during the last
stage of the optical transient (see Figure~\ref{fig_naked_power_lastpeak}), may be
interpreted as a Lense-Thirring precession or nutation
\cite{hartle,reynoso}. As the ejecta wobbles, the line of sight crosses
different parts of it with optical emission intensity varying by $\sim$10\%.
The non-detection of such a periodic feature in gamma-ray light curve may be
due to significant stochastic component on the similar time scale \cite{margutti}.

As a result of our wide-field high temporal resolution observations of
Naked-Eye Burst for the first time the fast (from one second till ten seconds)
optical prompt emission variability was discovered -- and this is the fastest
optically variable source seen on cosmological distances. The comparison of the
burst variability in optical and gamma-ray range for the first time definitely
revealed the direct connection of intensity variations with the activity of
central engine, which is, supposedly, a newborn stellar-mass black hole
accreting from massive hyperaccreting disk.

\bibliography{naked}

\bibliographystyle{Science}

% Following is a new environment, {scilastnote}, that's defined in the
% preamble and that allows authors to add a reference at the end of the
% list that's not signaled in the text; such references are used in
% *Science* for acknowledgments of funding, help, etc.

\begin{scilastnote}
\item This work was supported by the Bologna University Progetti Pluriennali
  2003, by grants of CRDF (No. RP1-2394-MO-02), RFBR (No. 04-02-17555 and
  06-02-08313), INTAS (04-78-7366), and by the Presidium of the Russian Academy
  of Sciences Program. We thank Emilio Molinari, Stefano Covino and Cristiano
  Guidorzi for technical help organizing TORTORA observations and for useful
  discussions of the results.
\end{scilastnote}

% For your review copy (i.e., the file you initially send in for
% evaluation), you can use the {figure} environment and the
% \includegraphics command to stream your figures into the text, placing
% all figures at the end.  For the final, revised manuscript for
% acceptance and production, however, PostScript or other graphics
% should not be streamed into your compliled file.  Instead, set
% captions as simple paragraphs (with a \noindent tag), setting them
% off from the rest of the text with a \clearpage as shown  below, and
% submit figures as separate files according to the Art Department's
% instructions.

% \clearpage

% \noindent {\bf Fig. 1.} Please do not use figure environments to set
% up your figures in the final (post-peer-review) draft, do not include graphics in your
% source code, and do not cite figures in the text using \LaTeX\
% \verb+\ref+ commands.  Instead, simply refer to the figure numbers in
% the text per {\it Science\/} style, and include the list of captions at
% the end of the document, coded as ordinary paragraphs as shown in the
% \texttt{scifile.tex} template file.  Your actual figure files should
% be submitted separately.

\end{document}